\documentclass{pasa}%

\usepackage{graphicx}

\title[Anisotropic winds in binaries]{Conditions for accretion disc formation and observability of wind-accreting X-ray binaries}

\author[R. Hirai \& I. Mandel]{Ryosuke Hirai$^{1,2}$, Ilya Mandel$^{1,2,3}$
\affil{$^1$OzGrav: Australian Research Council Centre of Excellence for Gravitational Wave Discovery, Clayton, VIC 3800, Australia}%
\affil{$^2$Monash Centre for Astrophysics, School of Physics and Astronomy, Monash University, Clayton, Victoria 3800, Australia}
\affil{$^3$Institute of Gravitational Wave Astronomy and School of Physics and Astronomy, University of Birmingham, Edgbaston, Birmingham B15 2TT, United Kingdom}
}%

\jid{PASA}
\doi{10.1017/pas.\the\year.xxx}
\jyear{\the\year}

\usepackage{aas_macros}
\usepackage{hyperref} 
\hypersetup{colorlinks,citecolor=blue,linkcolor=blue,urlcolor=blue}

\usepackage{bm}
\usepackage{acronym}
\usepackage{gensymb}

\newcommand{\msun}{\mathrm{M}_\odot}

\acrodef{BH}{black hole}
\acrodef{NS}{neutron star}
\acrodef{HMXB}{high-mass X-ray binary}
\acrodefplural{HMXB}[HMXBs]{high-mass X-ray binaries}
\acrodef{LMXB}{low-mass X-ray binary}
\acrodefplural{LMXB}[LMXBs]{low-mass X-ray binaries}
\acrodef{ISCO}{innermost stable circular orbit}

\usepackage{color}

\begin{document}

\begin{frontmatter}
\maketitle

\begin{abstract}
We explore the effect of anisotropic wind driving on the properties of accretion onto black holes in close binaries. We specifically focus on line-driven winds, which are common in high-mass X-ray binaries. In close binary systems, the tidal force from the companion star can modify the wind structure in two different ways. One is the reduction of wind terminal velocity due to the weaker effective surface gravity. The other is the reduction in mass flux due to gravity darkening. We incorporate these effects into the so-called CAK theory in a simple way and investigate the wind flow around the accretor on the orbital scale. We find that a focused accretion stream is naturally formed when the Roche lobe filling factor is $\gtrsim0.8$--0.9, analogous to that of wind Roche lobe overflow, but only when the velocity reduction is taken into account. The formation of a stream is necessary to bring in sufficient angular momentum to form an accretion disc around the black hole. Gravity darkening effects reduce the amount of accreted angular momentum, but not enough to prevent the formation of a disc. Based on these results, we expect there to be a discrete step in the observability of high-mass X-ray binaries depending on whether the donor Roche lobe filling factor is below or above $\sim$0.8--0.9.
\end{abstract}

\end{frontmatter}

\section{Introduction}\label{sec:introduction}

X-ray binaries are binary systems that consist of a regular star and a compact object, which can be a \ac{NS} or \ac{BH}. They are usually classified into two groups: \acp{HMXB} and \acp{LMXB}, depending on the mass of the donor star. \acp{HMXB}, in particular, can be important waystations for forming binary compact objects that eventually become gravitational wave sources \citep[e.g.][]{van19}. More than a hundred \acp{HMXB} have been identified so far, where $\sim2/3$ of them are known to host Be stars and the remaining $\sim1/3$ host O- or early-B supergiant stars \citep[]{liu06}. Among the supergiant \acp{HMXB}, there are only four known systems hosting \acp{BH} (BH-HMXBs). Observations of \acp{HMXB} have been critical in developing our understanding of massive star evolution, accretion physics and general relativity (see \citealt{mot21} for a review on \ac{BH} X-ray binaries).

One of the main differences between \acp{LMXB} and \acp{HMXB} is the mode of accretion onto the compact object. While \acp{LMXB} accrete through Roche lobe overflow, \acp{HMXB} accrete part of the stellar wind emitted from the donor. The overall framework of wind accretion can be well described by the classical Bondi-Hoyle-Lyttleton model \citep[BHL;][]{hoy39,bon44}. However, there are also known features in some observed BH-HMXBs such as Cygnus X-1, which cannot be explained by quasi-spherical accretion as in the BHL model.

Cygnus X-1 (Cyg X-1), which was discovered in the 1960s \citep[]{bow65,bol72}, is one of the four known supergiant BH-HMXBs and also one of the very few persistent X-ray sources. The \ac{BH} mass has recently been estimated to be $\sim21~\msun$, orbited by an O-type star with a mass of $\sim40~\msun$ \citep[][]{mil21}. The star is extremely close to filling its Roche lobe, with a relatively short orbital period of 5.6~d \citep[]{web72}. The \ac{BH} is rapidly spinning, with a dimensionless spin parameter of $\chi\gtrsim0.9$ \citep[][see, however \citealp{mil09,Kawano:2017}]{gou11,gou14,dur16,Zhao:2021}. Like most other X-ray binaries, Cyg X-1 frequently transitions between the low-hard state and high-soft state. The existence of a focused accretion stream has been identified in the low-hard and intermediate states \citep[]{mil05,pou08,han09}, indicating a clear deviation from the simple BHL accretion picture.

The formation of a focused stream in close binaries has in fact been suggested by \citet{fri82}. Their model builds upon the theory constructed by \citet{cas75} for line-driven winds. \citet{fri82} account for the change in the effective gravitational force due to the presence of a binary companion, which causes the wind mass flux to increase in the direction of the secondary. However, there are some simplifications made in the model that may lead to an overestimate of the effect such as ignoring the Coriolis force, assuming axial symmetry etc.

Some hydrodynamical simulations of wind accretion in X-ray binaries have reproduced the focused flows \citep[]{blo90,blo91,nag04,jah05,had12,cec15,elm19}. Each simulation adopts different ways to drive the wind in the simulations, sometimes hidden in the details of the boundary conditions. Although it is challenging to quantitatively compare these models, it seems that the necessary condition for stream formation is that the companion star is close to filling its Roche lobe. Indeed, all wind-fed \acp{HMXB} with confirmed BHs have rather high Roche lobe filling factors; Cyg X-1 \citep[0.997;][]{mil21}, LMC X-1 \citep[0.971;][]{oro09}, M33 X-7 \citep[0.899;][]{oro07}\footnote{Note that our definition of the Roche lobe filling factor (see section \ref{sec:model}) is different from the cited papers and thus these values are higher than their quoted values \citep[see][]{Neijssel:2021}.}.  Therefore, there may be a strong connection between stream formation and the observability of \acp{HMXB}.

In this paper, we investigate the necessary conditions for stream formation and the associated accretion disc formation in wind-fed X-ray binaries. We mainly focus on the influence of anisotropic driving of the wind due to two different effects. One is the change in the wind velocity structure due to the aspherical gravitational field in binaries. This is similar to the effect explored in \citet{fri82}, but we lift some of the assumptions made in their model. The other effect is the gravity darkening effect. The donor star should have an inhomogeneous radiative flux at the surface due to tidal effects. We expect this to influence the mass flux along each streamline, significantly impacting the amount of mass flowing towards the companion. We construct a model to incorporate these effects in a simple way and explore how each aspect impacts the amount of mass and angular momentum that accretes onto the compact object. 

This paper is structured as follows. We first revisit the standard theory for line-driven winds and explain our approach to incorporate the two extra effects in Section~\ref{sec:formulation}. We then outline our numerical method and models in Section~\ref{sec:model}. The results are presented in Section~\ref{sec:results} and discussed in Section~\ref{sec:discussion}. We finally summarize our conclusions in Section~\ref{sec:conclusion}.

\section{Formulation of line-driving in close binary systems}
\label{sec:formulation}

\subsection{Line-driven wind theory for single non-rotating stars}
Here we briefly review the theory of line-driven winds for single non-rotating stars \citep[see Chapter 8 of][]{lam99}. 
In the atmospheres of hot luminous stars, the ions can have many narrow absorption lines. These lines can efficiently capture part of the ultraviolet radiation from the star and then symmetrically re-emit, so that the ions gain a net outwards momentum from the incoming photons. Due to the strong coupling between the ions and the surrounding gas, the atmosphere is accelerated outwards as a stellar wind. The acceleration in turn enables the lines to capture radiation over different frequencies via Doppler shifting: gas moves at greater velocities further away from the star, so can capture redder photons that passed through gas at smaller radii. \citet{cas75} (hereafter CAK) established that the acceleration due to spectral lines can be approximated as a power law\footnote{See however \citet{lat21} for recent calculations that suggest a deviation from a power law.}
\begin{equation}
 g_\mathrm{L}=\frac{\sigma_eL}{4\pi r^2c}kt^{-\alpha},
\end{equation}
where $\sigma_e$ is a reference value for the electron scattering opacity, $L$ the luminosity of the star, $r$ is the distance from the centre of the star, and $c$ the speed of light.  The dimensionless optical depth parameter $t$ is given by
\begin{equation}
 t\equiv\sigma_ev_\mathrm{th}\rho\left(\frac{dv}{dr}\right)^{-1},
\end{equation}
where $v_\mathrm{th}$ is the mean thermal velocity of the protons in the wind, $\rho$ the density and $v$ the velocity of the wind. The so-called force multiplier parameters $k$ and $\alpha$ are tabulated for various types of stars \citep[e.g.][]{abb82,shi94,lat21}. For hot stars with effective temperatures $3.8\lesssim\log{(T_\mathrm{eff}/K)}\lesssim4.7$, these (correlated) parameters fall in the ranges $0.1\lesssim k\lesssim1$ and $0.45\lesssim\alpha\lesssim0.7$.

\subsubsection{Point source limit} \label{sec:pointsource}

Given the expression for the line acceleration, the momentum equation can be written in the point source limit as
\begin{equation}
 v\frac{dv}{dr}=-\frac{GM}{r^2}-\frac{1}{\rho}\frac{dp}{dr}+g_e+g_\mathrm{L},
\end{equation}
where $G$ is the gravitational constant, $M$ the stellar mass, $p$ is pressure and 
\begin{equation}
 g_e=\frac{GM}{r^2}\Gamma
\end{equation}
is the continuum acceleration. Here we have defined $\Gamma\equiv\sigma_eL/(4\pi cGM)$ which is the Eddington factor. By using the mass continuity equation ($\rho=|\dot{M}|/(4\pi r^2v)$) and some simple manipulation we get
\begin{equation}
 \left(1-\frac{c_s^2}{v^2}\right)r^2v\frac{dv}{dr}=-GM(1-\Gamma)+2c_s^2r+C\left(r^2v \frac{dv}{dr}\right)^\alpha,\label{eq:eom_pointsource}
\end{equation}
where $c_s\equiv\sqrt{dp/d\rho}$ is the isothermal sound speed and $C$ is a constant that contains all the model parameters 
\begin{equation}
 C\equiv\frac{\sigma_eLk}{4\pi c}\left(\frac{\sigma_ev_\mathrm{th}|\dot{M}|}{4\pi}\right)^{-\alpha}.\label{eq:Cdef}
\end{equation}
We hereafter ignore the pressure gradient ($c_s\rightarrow0$), which is known to only make a minor difference in the velocity distribution. Requiring that Eq.~(\ref{eq:eom_pointsource}) has a unique solution, the velocity can be solved for analytically 
\begin{equation}
 v(r)=\sqrt{\frac{\alpha}{1-\alpha}}v_\mathrm{esc}\left(1-\frac{R}{r}\right)^{0.5},\label{eq:vel_pswind}
\end{equation}
where $v_\mathrm{esc}\equiv\sqrt{2GM(1-\Gamma)/R}$ is the escape velocity of the star. The parameter $C$ cannot be arbitrarily chosen but has to take a specific form
\begin{equation}
 C=\frac{1}{\alpha}\left(\frac{\alpha}{1-\alpha}GM(1-\Gamma)\right)^{1-\alpha}.
\end{equation}
By equating this with the definition of $C$ (Eq.~(\ref{eq:Cdef})), the mass-loss rate $\dot{M}$ can be uniquely obtained given all the other stellar parameters.
We can see from Eq.~(\ref{eq:vel_pswind}) that the wind has a terminal velocity 
\begin{equation}
 v_\infty=\sqrt{\frac{\alpha}{1-\alpha}}v_\mathrm{esc},
\end{equation}
which is very close to the escape velocity of the star in the typical range of $\alpha$ ($\alpha\sim0.45$--0.7). Eq.~(\ref{eq:vel_pswind}) corresponds to a so-called $\beta$-law which takes the form 
\begin{equation}
v=v_\infty\left(1-\frac{R}{r}\right)^\beta,
\end{equation}
 with $\beta=0.5$.

\subsubsection{Finite disc correction} \label{sec:finitedisc}
In reality, stars are not point sources and have finite sizes. The wind material can feel radiative forces from a wide range of angles when it is still close to the surface of the star. The net outwards momentum can thus be reduced close to the surface, as many photons approach the wind material from a non-radial angle. This effect can be accounted for by simply multiplying the line-driven acceleration by
\begin{equation}
 K_\textsc{fdcf}=\frac{(1+\sigma)^{\alpha+1}-(1+\sigma\mu_*^2)^{\alpha+1}}{(1-\mu_*^2)(\alpha+1)\sigma(1+\sigma)^\alpha},
\end{equation}
where $\sigma\equiv d\ln v/d\ln r-1$, and $\mu_*\equiv\sqrt{1-(R/r)^2}$. This gives $K_\textsc{fdcf}=1/(1+\alpha)$ at $r=R$ and $K_\textsc{fdcf}\rightarrow1$ for $r\rightarrow\infty$. So the momentum equation with the finite disc correction becomes 
\begin{equation}
 r^2v\frac{dv}{dr}=-GM(1-\Gamma)+CK_\textsc{fdcf}\left(r^2v \frac{dv}{dr}\right)^\alpha.\label{eq:eom_finitedisc}
\end{equation}
Although this cannot be solved analytically, it is known that the solution to this equation can be well fitted by a $\beta$-law of
\begin{equation}
 v(r)=2.5\cdot \frac{\alpha}{1-\alpha}v_\mathrm{esc}\left(1-\frac{R}{r}\right)^{0.7},\label{eq:vwind_fd_beta}
\end{equation}
based on both analytical \citep[]{pau86} and numerical \citep[]{mul08} studies.
Notice that the terminal velocity is a few times faster than that of the point source limit. The factor 2.5 is consistent with both observations \citep[]{lam95} and numerical simulations \citep[]{vin99} for stars with temperatures $T\gtrsim21,000$~K. This velocity law can be approximately obtained by setting the constant $C$ to
\begin{equation}
 C=C_0\equiv\frac{1+\alpha}{\alpha}\left[\frac{\alpha}{1-\alpha}GM(1-\Gamma)\right]^{1-\alpha}\label{eq:C_standard}
\end{equation}
in Eq.~(\ref{eq:eom_finitedisc}) and integrating from the surface of the star outwards.

\subsection{Velocity distribution for rotating stars}\label{sec:rotation}

When the star is rotating, an extra term accounting for the centrifugal force enters the momentum equation, based on the assumption that the material in the wind conserves the specific angular momentum of the surface of the star. This alters the velocity distribution of the wind. Numerical investigations show that the terminal velocity decreases with increasing rotational velocity without significantly affecting the power $\beta$ \citep[][]{cas79,abb80,fri86,cur04,owo06,mad07}.

\begin{figure}
 \centering
 \includegraphics[width=\linewidth]{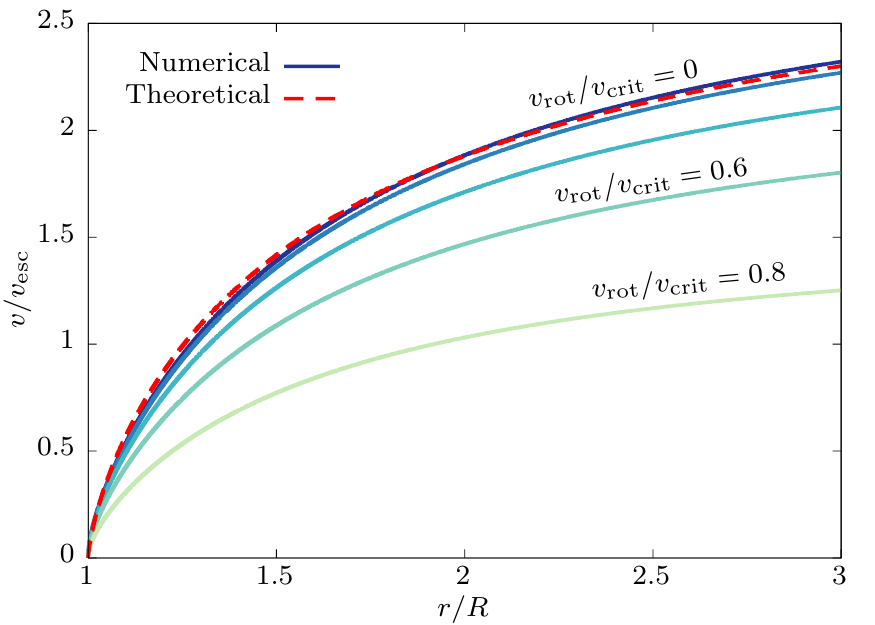}
 \caption{Wind velocity distribution for stars with different rotational velocities. Model parameters  $\alpha=0.55$ and $\Gamma=0.1$ are used for this calculation. Solid curves are the numerically computed values for rotational velocities $v_\mathrm{rot}/v_\mathrm{crit}=0, 0.2, 0.4, 0.6, 0.8$ and the red dashed curve shows the $\beta$-law presented in Eq.~(\ref{eq:vwind_fd_beta}).\label{fig:vwind_rotation}}
\end{figure}

The wind velocity decreases because the escape velocity is effectively reduced due to the centrifugal force. With this in mind, we apply an extra factor in the constant $C$ to account for the reduction in effective gravity 
\begin{equation}
 C=C_\mathrm{rot}\equiv C_0\left[1-\left(\frac{v_\mathrm{rot}}{v_\mathrm{crit}}\right)^2\right]^{1-\alpha},\label{eq:C_rot}
\end{equation}
to see if we can reproduce the wind profile for rotating stars. Here $v_\mathrm{rot}$ is the equatorial rotational velocity of the star and $v_\mathrm{crit}\equiv\sqrt{GM(1-\Gamma)/R}$ is the critical rotational velocity. In Figure~\ref{fig:vwind_rotation}, we show the velocity distribution obtained by integrating Eq.~(\ref{eq:eom_finitedisc}) with the constant $C$ as in Eq.~(\ref{eq:C_rot}). The non-rotating case is in good agreement with the theoretical prediction expressed by Eq.~(\ref{eq:vwind_fd_beta}). As the rotational velocity increases, the wind velocity decreases while preserving the overall $\beta$-law shape, which is consistent with previous findings \citep[]{fri86}.

\begin{figure}
 \centering
 \includegraphics[width=\linewidth]{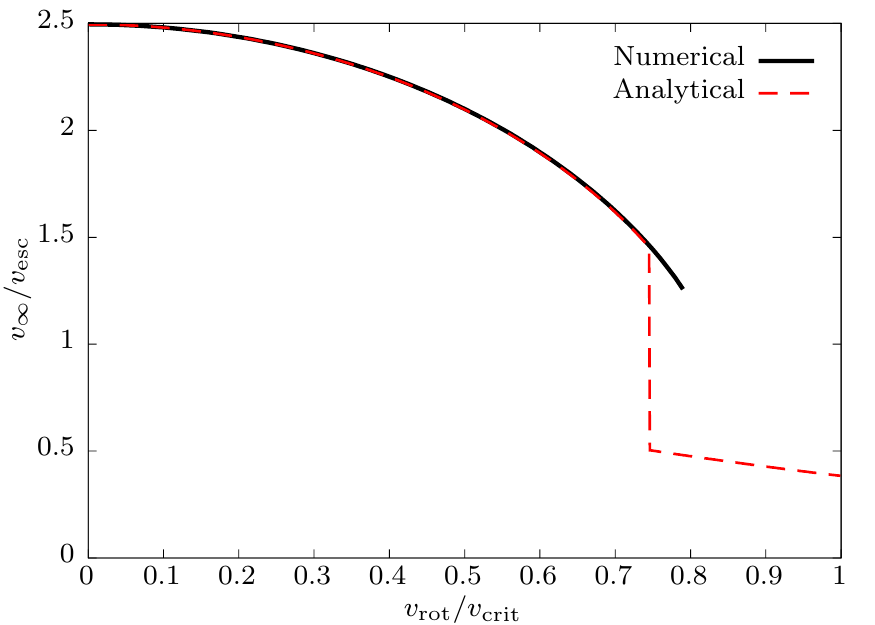}
 \caption{Terminal wind velocity as a function of rotational velocity. The black solid curve shows the numerical value while the red dashed curve shows the analytical value following the analysis of \citet{owo06}. The acceleration parameter used for this figure is $\alpha=0.5$.\label{fig:vterm_vrot}}
\end{figure}

We plot the terminal wind velocity against the rotational velocity in Figure~\ref{fig:vterm_vrot}. Our numerically obtained values are in almost perfect agreement with the analytical values based on a simple ``nozzle analysis'' presented in \citet{owo06}. The nozzle analysis uses the fact that the parameter $C$ is proportional to the mass flux ($C\propto|\dot{M}|^{-\alpha}$). The maximum allowed mass flux is determined by the condition that the velocity profile should be monotonically increasing outwards. The integration fails in our models with rotation velocities faster than $v_\mathrm{rot}/v_\mathrm{crit}\gtrsim0.8$, which is roughly consistent with the critical rotation derived from the nozzle analysis as indicated by the jump in the red dashed curve \citep[see also][]{ara18}. The perfect agreement between our model and the nozzle analysis at $v_\mathrm{rot}/v_\mathrm{crit}\lesssim0.75$ implies that setting the constant $C$ to Eq.~(\ref{eq:C_rot}) is equivalent to restricting the mass flux to the ``nozzle'' value. 

\subsection{Equation of motion in a binary system}

When the wind-emitting star is in an X-ray binary system, the gravitational force of the compact object should also be included in the equation of motion of the wind. The full equation of motion in the corotating frame becomes 
\begin{align}
 \frac{d\bm{v}}{dt}=&-\frac{GM_1(1-\Gamma)}{|\bm{r}-\bm{r}_1|^2}\bm{n}_1-\frac{GM_2}{|\bm{r}-\bm{r}_2|^2}\bm{n}_2\nonumber\\
 &+CK_\textsc{fdcf}\left(v_1|\bm{r}-\bm{r}_1|^2 \bm{n}_1\cdot\nabla\left[\bm{n}_1\cdot\bm{v}\right]\right)^\alpha \frac{\bm{n}_1}{|\bm{r}-\bm{r}_1|^2}\nonumber\\
 &-\bm{\Omega}\times(\bm{\Omega}\times\bm{r})-2\bm{\Omega}\times\bm{v},\label{eq:eom_full}
\end{align}
where $M_i, \bm{r}_i$, $\bm{n}_i(\bm{r})\equiv(\bm{r}-\bm{r}_i)/|\bm{r}-\bm{r}_i|$ $(i=1, 2)$ are respectively the mass, position vector and normal vector pointing from the $i$-th star and $\bm{\Omega}=\sqrt{G(M_1+M_2)/a^3}$ is the orbital angular velocity where $a\equiv|\bm{r}_1-\bm{r}_2|$ is the orbital separation. The first term includes the gravitational and continuum acceleration from the donor (primary), the second term is the gravitational acceleration from the accretor (secondary), the third term expresses the line-driven acceleration from the primary star, the fourth and fifth terms correspond to the centrifugal and Coriolis force. We will discuss the value of the parameter $C$ in the following section.

\subsection{Tides and gravity darkening}
Here we are focusing on the case where the secondary star is a compact object (\ac{BH} or \ac{NS}) and therefore the primary star is the only wind emitter. In a close binary system on a circular orbit, the donor star can feel a strong tidal force from the companion and becomes synchronized with the binary orbit. The star would then adjust its shape such that it fills the equipotential surface of the Roche potential. This distortion of the stellar shape along with the tidal force from the secondary causes the radial acceleration on the surface of the star to vary drastically depending on the position. Since the terminal velocity of winds are strongly related to the surface effective gravity (see Section~\ref{sec:rotation}), the wind velocity field could be highly asymmetric. We approximately account for this effect by choosing the value of $C$ proportional to the surface effective gravity as
\begin{equation}
  C=C_\mathrm{bin}\equiv C_0\left(\frac{|\bm{g}_0|}{|\bm{g}_\mathrm{pole}|}\right)^{1-\alpha},\label{eq:C_binary}
\end{equation}
where $\bm{g}_0$ is the radial acceleration at the base of a wind streamline (surface of the star) and $\bm{g}_\mathrm{pole}$ is the radial acceleration at the pole of the star. Here the acceleration includes gravity from both stars and the centrifugal force. This expression is equivalent to Eq.~(\ref{eq:C_rot}) when applied to single rotating stars. In binaries, this will lead to a non-axisymmetrical wind distribution, with lower wind velocities in the direction towards and away from the accretor.

In addition to determining the wind velocity distribution, the effective surface gravity is expected to change the surface radiative flux of the star via gravity darkening\footnote{The conventional name is actually misleading. It is the reduction of gravity that causes the darkening, so strictly speaking, it should be levitation darkening.}. Based on von Zeipel's theorem \citep[]{von24}, the radiative flux $F$ at the surface of the star scales as
\begin{equation}
 F\propto T_\mathrm{eff}^4\propto |\bm{g}_0|^{\beta_1},
\end{equation}
where $T_\mathrm{eff}$ is the effective temperature of the star and $\beta_1$ is the gravity darkening exponent. The exponent is $\beta_1=1$ in the original von Zeipel theorem which was derived for radiative stars, but can take much lower values for convective stars \citep[$\beta_1\sim0.3$--0.4;][]{luc67,cla12}. In rotating stars, this leads to a darkening of the equator due to the lower effective gravity, which has actually been observed for the case of Altair \citep[]{des05}.

\begin{figure}
 \centering
 \includegraphics[width=\linewidth]{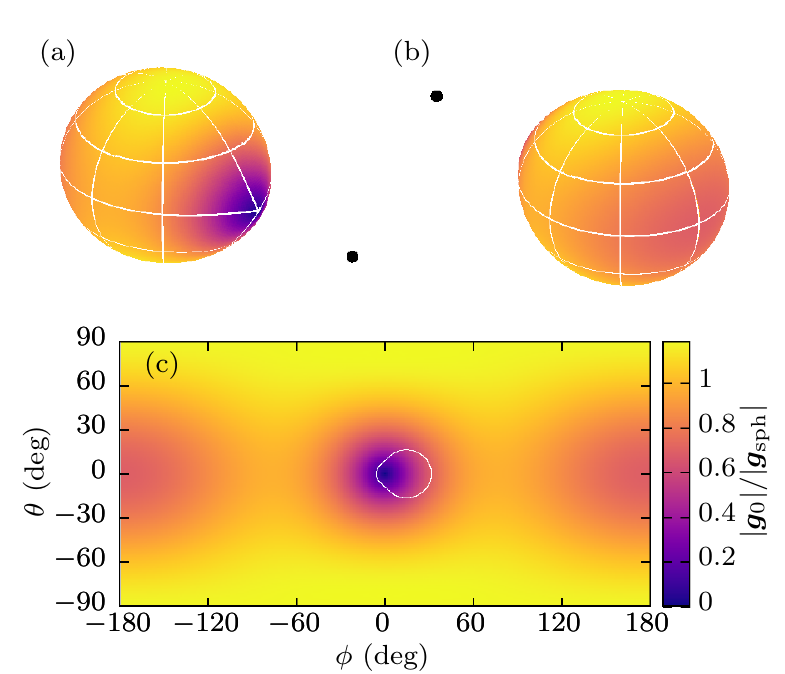}
 \caption{Illustration of the effective gravity distribution over the surface of the star in a tidally locked equal-mass binary with a Roche lobe filling factor of 1. Panels (a) and (b): 3D shape of the tidally distorted star from two different angles. The surface is coloured by the surface acceleration. The black dots indicate the position of the accretor. Panel (c): Mercator projection of the surface acceleration. The coordinate origin points in the direction of the accretor star. The acceleration is normalised by the surface acceleration of a spherical star with the same volume. The wind originating from the region surrounded by the white line intersects with the accretion radius of the companion when the velocity correction is on (see section \ref{sec:models}).\label{fig:surface_grav}}
\end{figure}

Due to the tidal force, the surface flux distribution on a star in a close binary system can be even more asymmetric than rotating single stars. For example, for a fully Roche lobe filling star, the effective gravity at the first Lagrangian point is zero, meaning that the radiative flux should also be zero at that point. Overall, such a star will have a vanishing flux at the point facing toward the accretor, a greatly reduced flux away from the accretor, slightly darkened sides and bright poles \citep[]{whi12,had12,cec15}. This effect appears as ``ellipsoidal variations'' in the optical light curves of X-ray binary donors \citep[e.g.][]{avn75}. We illustrate the surface effective gravity map in Figure~\ref{fig:surface_grav}. It is clear that there are two dark patches, but the patch facing towards the accretor is much darker than the opposite side. The acceleration is normalised by the surface acceleration of a spherical star with the same mass and volume as the distorted star ($\bm{g}_\mathrm{sph}$). Notice that the acceleration at the poles slightly exceeds that of a spherical star, as the tidal force acts to flatten the star in the vertical direction.

Since the stellar winds of massive stars are mainly driven by radiation, gravity darkening will cause the wind mass flux to vary over the surface of the star. This has been thoroughly studied theoretically for the case of rotating stars where the polar mass flux is expected to be enhanced over the equatorial mass flux \citep[]{cra95,owo96,owo97,owo98,gag19}. Although gravity darkening has been observed for some stars, observational evidence for its influence on the wind driving is still ambiguous. However, such polar-enhanced winds are known for some massive stars such as $\eta$ Carinae \citep[]{smi03}. The rotation of the star has not been measured, but given all the other indirect evidence supporting a stellar merger origin for $\eta$ Carinae \citep[e.g.][]{smi18,RH21}, it is natural to associate the latitudinal dependence with rapid rotation. We therefore expect that the asymmetric gravity darkening in close binaries could also lead to highly asymmetric mass flux. We take this into account by assuming 
\begin{equation}
 \frac{\dot{M}}{\dot{M}_\mathrm{pole}}=\left(\frac{|\bm{g}_0|}{|\bm{g}_\mathrm{pole}|}\right)^{\beta_1},\label{eq:mass-loss_rate}
\end{equation}
where $\dot{M}$ is the local mass flux and $\dot{M}_\mathrm{pole}$ is the mass flux at the pole. 

We ignore any variations of $\alpha$ over the surface, although this could also be an important factor as suggested in the context of B[e] stars via the so-called bi-stability mechanism, which is caused by the change in ionization below some threshold effective gravity \citep[]{lam91}.

\begin{figure}
 \centering
 \includegraphics[width=\linewidth]{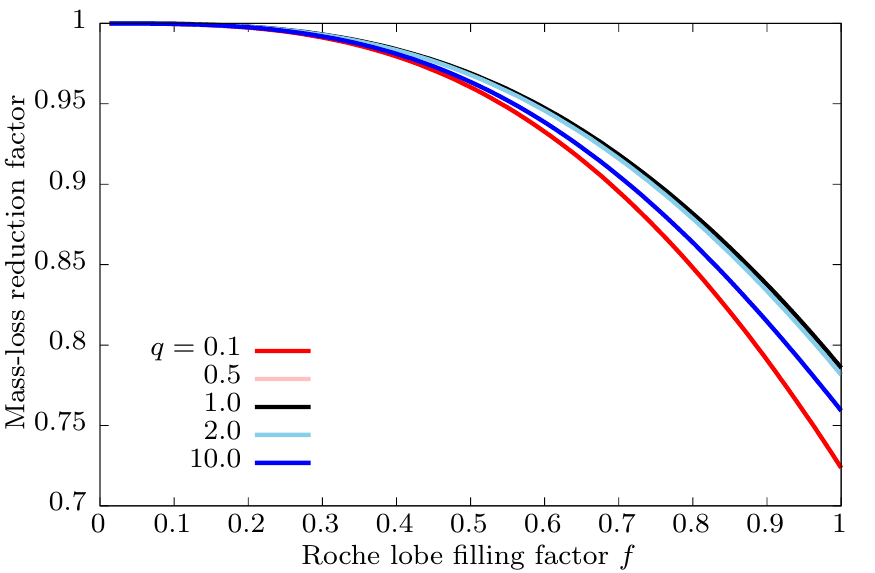}
 \caption{Total reduction of mass flux due to gravity darkening in binaries. Different colours show results for different mass ratios, which is defined as $q\equiv M_2/M_2$. All curves were computed assuming maximal gravity darkening ($\beta_1=1$). The $q=0.5$ curve is almost identical to the $q=2.0$ curve and is therefore not visible. \label{fig:tot_mdot}}
\end{figure}

Figure~\ref{fig:tot_mdot} displays the total mass flux from a gravity darkened star ($\beta_1=1$) compared to that of a non-darkened case ($\beta_1=0$). For an equal mass binary ($q=1$), the total mass flux can be reduced by up to $\sim20\%$ when it is close to filling its Roche lobe. The reduction is slightly larger for unequal mass binaries, but the mass ratio dependence is minor.

\subsection{Caveats} \label{sec:caveats}

We made several simplifying assumptions in our formulation of the wind acceleration. First, we have ignored the pressure gradient in the wind acceleration, which can slightly raise the terminal velocity. Multiple scattering of photons have been ignored, which could lead to enhanced winds in optically thick cases \citep[e.g.][]{fri83,abb85}. While multiple scattering can be important for stars like Wolf-Rayet stars, it is expected to be minor in the optically thin winds of O/B stars. We have assumed the radiative forces from the donor are always radial. This is not strictly true when the star suffers gravity darkening and there are flux variations over the surface \citep[]{gay00}. However, the non-radial component of the radiative force is at most $\lesssim10~\%$ and rapidly decreases with distance from the star \citep[]{cec15}. There also could be limb darkening effects as the stars do not have sharp surfaces \citep[]{cra95}. 

Another key assumption is that we use the same  Eddington factor and acceleration parameter $\alpha$ over the entire streamline, using the value at its launching point. This can be incorrect around the locations where there are significant flux and temperature variations due to gravity darkening, or when the streamlines are bent significantly by the Coriolis force. The former can reduce the effect of gravity darkening, because the wind material can see brighter parts of the star as it travels out, even if it was launched at much darker patches. Also, we have ignored any variations of $\alpha$ over the surface of the star. For example, the bi-stability mechanism may boost the mass flux at the darker areas, counteracting the gravity darkening effect we apply here \citep[]{lam91,owo97,owo98,gag19}.

Additionally, we have assumed a stationary wind velocity and density distribution. Line-driven winds are subject to instabilities, which may cause large variabilities leading to X-ray emission\footnote{These X-rays generated within the wind can serve as an additional heating source to slightly raise the terminal wind velocity \citep[]{san18}.} (see \citealt{owo94} for a theoretical review on wind instability). As a result, the wind can become very clumpy, which can in principle cause stochastic accretion patterns \citep{osk12}. However, the inhomogeneities on the orbital scale can be wiped out by hydrodynamical interactions before reaching the X-ray emitting region \citep[]{elm18}. Indeed, the observed X-ray variability is much smaller than what is predicted by simply translating the orbital scale stochasticity to X-rays \citep[]{osk12}. Also, the time-averaged wind profiles usually follow the stationary solution \citep[e.g.][]{owo88}, so the actual effect of wind clumping may be limited in the long term.

One major effect we have ignored in this study is the X-ray feedback from the accretor. The ionization structure of the wind material can be altered by the irradiation, strongly influencing the wind driving \citep[]{hat77,mac82,mas84,ste90,san18}. When the metals become photoionized by the X-rays, the UV absorption lines are significantly weakened. The radiative flux from the donor cannot push on the wind any more, reducing the wind terminal velocity. The consequences of diminishing acceleration strongly depend on how far the ionization front reaches\footnote{Analogous to a Str\"omgren sphere.}, which in turn depends on the luminosity of the X-ray feedback. If the feedback is relatively weak, the photoionization only occurs in a region where the wind has already accelerated sufficiently, and the effect of ionization acts to decrease the terminal velocity. This will increase the accretion radius and therefore raises the total mass accretion rate \citep[e.g.][]{kar14,cec15}. In the presence of strong feedback where the ionization front reaches the surface of the donor, it can completely shut off the wind from the hemisphere facing the accretor \citep[]{blo94,krt12,cec15,krt16,krt18}. Such a situation cannot be sustained because the X-ray feedback will eventually die out as the accretor exhausts the matter that was powering the strong X-rays in the first place. This means that HMXBs may be in a self-regulated state where the X-ray luminosity varies around the critical value for wind inhibition \citep[]{krt12,krt18}. \citet{krt16} show that many X-ray binaries including Cyg X-1 have X-ray luminosities close to the wind inhibition limit.  This effect may be weakened due to wind clumping \citep[]{osk12}. Although the X-ray irradiation is indeed a critical ingredient to understand the true wind accretion process, we shall leave quantitative investigation for future work and only briefly discuss the qualitative effects in this paper.

\section{Method \& models}\label{sec:model}
In this paper, we explore the effects of rotation and tides on winds in tight X-ray binaries. We especially focus on the amount of mass and angular momentum that could accrete onto the compact object. To efficiently explore the wide parameter space and for simplicity, we take a ballistic approach where we integrate the equation of motion (Eq.~(\ref{eq:eom_full})) ignoring any interactions between the streamlines. A full understanding of the accretion process onto the companion compact object would require a 3D hydrodynamical simulation (e.g. \citealp{nag04,jah05,had12,cec15,elm18,elm19,mac20,sch21}, see also \citealp{blo90,blo91,pal20} for 2D approaches). Here, we focus on the qualitative differences introduced by taking into account the effects of reduced effective surface gravity and the possible associated gravity darkening.

\subsection{Method}
Most of our methodology is similar to \citet{elm17} except for the asymmetrical effects on the wind acceleration, which is the main focus of this paper.

To model the tidally distorted donor surface, we compute the equipotential surface within the Roche potential
\begin{equation}
 \Phi(\bm{r})=-\frac{GM_1}{|\bm{r}-\bm{r}_1|}-\frac{GM_2}{|\bm{r}-\bm{r}_2|}-\frac12|\Omega\times\bm{r}|^2,\label{eq:roche_potential}
\end{equation}
assuming that the star is tidally locked to the orbit. Using the Roche potential implies point masses for the binary components, which is not a bad assumption given that massive main sequence stars have very centrally concentrated mass distributions. We characterize the size of the donor star with the Roche lobe filling factor defined as $f\equiv(V_*/V_\mathrm{RL})^{1/3}$, where $V_*$ is the volume of the star, assumed to fill a region bounded by an equipotential surface, and $V_\mathrm{RL}$ is the volume of the Roche lobe. Varying $f$ corresponds to either changing the stellar size with a fixed orbital separation, or changing the orbital separation for a fixed star.

From each point on the surface of the star, we integrate the equation of motion (\ref{eq:eom_full}) to obtain the velocity distribution of the wind. In order to calculate the amount of mass and angular momentum that accretes onto the companion, we draw a sphere around the compact object with a radius equal to the BHL accretion radius \citep[]{hoy39,bon44} expressed as
\begin{equation}
 r_\mathrm{acc}=\frac{2GM_2}{v_\bullet^2},
\end{equation}
where $v_\bullet$ is the wind velocity in the vicinity of the accreting object. For the value of $v_\bullet$, we use the vector sum of the orbital velocity and the radial wind velocity at a distance $a$ from a single rotating star with the same volume and angular velocity as the tidally distorted star. At a distance $r_\mathrm{acc}$ from the accretor, the wind is already significantly focused and hydrodynamical effects could become important. We consider an extended accretion radius $r_\mathrm{ext}>r_\mathrm{acc}$ to compare results at different scales.

\subsection{Models}\label{sec:models}
We use units of $G=a=M_1=1$ for all of our calculations. The mass ratio is defined as $q\equiv M_2/M_1$. We carry out calculations for a grid of binary models with different mass ratios $q=0.5, 1, 2$ and Roche lobe filling factors $f\in(0,1]$. We also vary the force multiplier parameter $\alpha=0.5, 0.6$ and Eddington factor $\Gamma=0.1, 0.3$ to compare results for stars with different surface properties.

For each binary model, we compute two different wind velocity distributions with different expressions for the constant $C$ when integrating the equation of motion: one with $C=C_0$ (Eq.~(\ref{eq:C_standard})) and one with $C=C_\mathrm{bin}$ (Eq.~(\ref{eq:C_binary})). The latter includes a reduction factor proportional to the lower effective surface gravity, and hereafter we will call this the velocity correction (VC). We also investigate the impact of gravity darkening (GD), by setting $\beta_1=0$ or 1 in Eq.~(\ref{eq:mass-loss_rate}). $\beta_1=1$ corresponds to a fully gravity darkened case and $\beta_1=0$ corresponds to a uniformly bright star. Note that in our formulation, the VC only influences the velocity distribution whereas GD only affects the mass flux along each streamline.

\section{Results}\label{sec:results}
\subsection{Streamline distribution}

\begin{figure*}
 \centering
 \includegraphics[width=\linewidth]{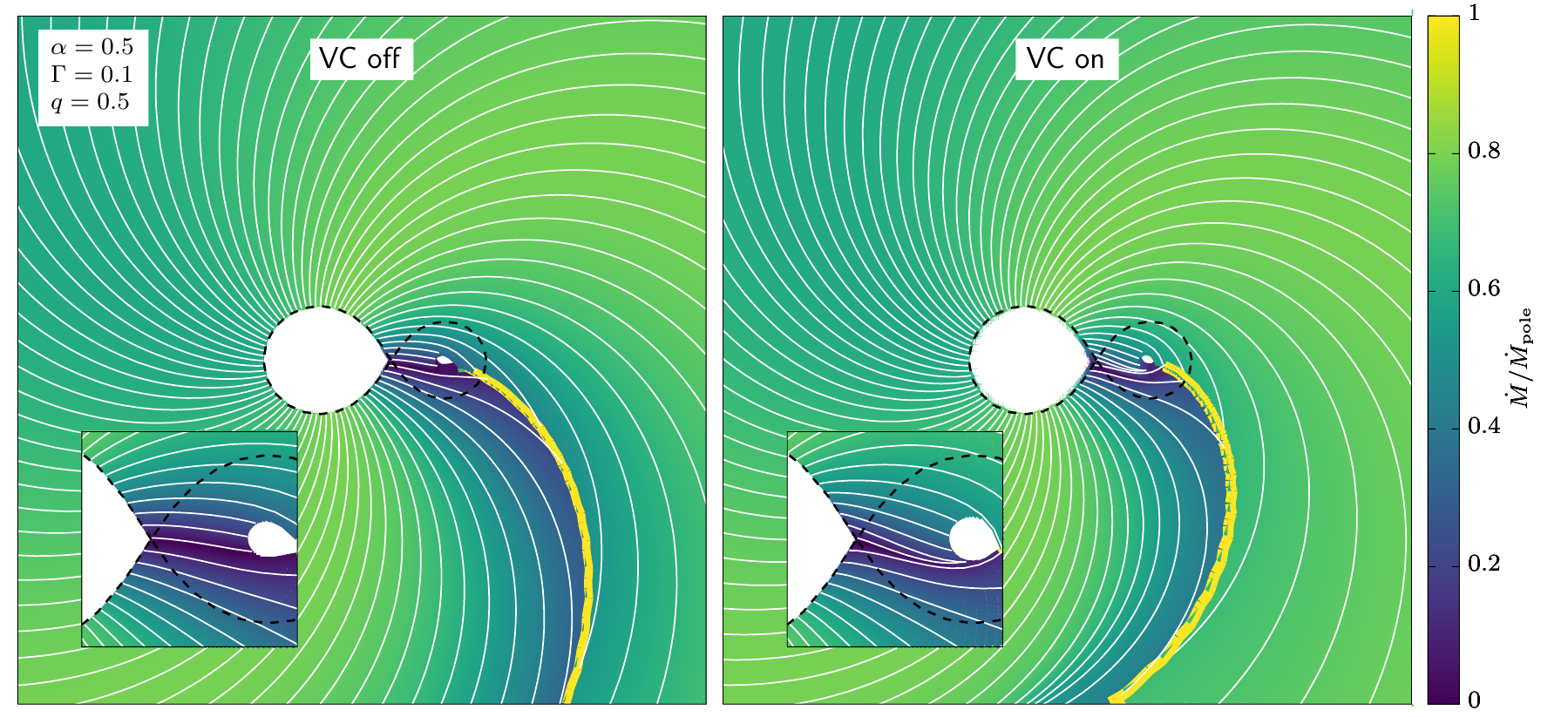}
 \caption{Wind streamlines in the equatorial plane (white curves). Model parameters are listed in the legend. The orbital angular momentum points out of the page. The background is coloured by the mass flux along each streamline normalized by the mass flux at the pole of the donor, assuming fully efficient GD ($\beta=1$). Black dashed curves indicate the Roche lobe. Yellow curves show the locus of momentum cancellation in the equatorial plane. Insets show zoomed-in images of the accretion stream.  Left and right panels compare the streamlines with and without the VC. \label{fig:streamlines_q05}}
\end{figure*}

Figure~\ref{fig:streamlines_q05} shows the wind streamline distribution for a binary model with $q=0.5, f=1$ and force parameters $\alpha=0.5, \Gamma=0.1$. The left panel does not include the VC and is equivalent to previous models in the literature \citep[e.g.][]{elm17}, whereas the right panel includes the VC. The most important feature is the formation of a focused stream towards the accretor in the right panel. The streamlines emanating from the vicinity of the first Lagrangian point are channeled as a stream, slightly diverted downwards due to the Coriolis force. This closely resembles the so-called ``wind Roche-lobe overflow'' \citep[]{nag04,pod07,moh07,moh10}.

For illustrative purposes, we truncated the streamlines at a position where the perpendicular component of the wind momentum balances that of another streamline (yellow curve). This is equivalent to assuming fully efficient cooling as the streamlines interact, giving us a rough idea of the shape of the dense tail that may form behind the accretor as seen in many hydrodynamical simulations \citep[]{blo90,blo91,nag04,jah05,had12}. Note that this is not an accurate representation since we only compute the momentum balance between streamlines on the equatorial plane, while in reality, streamlines from slightly higher and lower latitudes can be focused onto the plane and alter the morphology of the tail. Also, if the cooling is inefficient, a bow shock may form instead of a dense tail \citep[]{elm19,mac20,sch21}. Leaving the caveats aside, there is a noticeable difference in the shape of the tail that forms behind the accretor, which may be observable in high-mass X-ray binaries by looking at the orbital phase dependent X-ray absorption \citep[]{wen99,bal00}.

The background is coloured according to the mass flux along each streamline, for a fully gravity darkened model ($\beta_1=1$). Without GD ($\beta_1=0$), all streamlines have the same mass flux in our formulation. As expected, the mass flux from the vicinity of the first Lagrangian point is severely quenched by GD. The flux from the other side of the star is also quenched, although not as much as from the front side.

\subsection{Accreted mass}

\begin{figure}
 \centering
 \includegraphics[width=\linewidth]{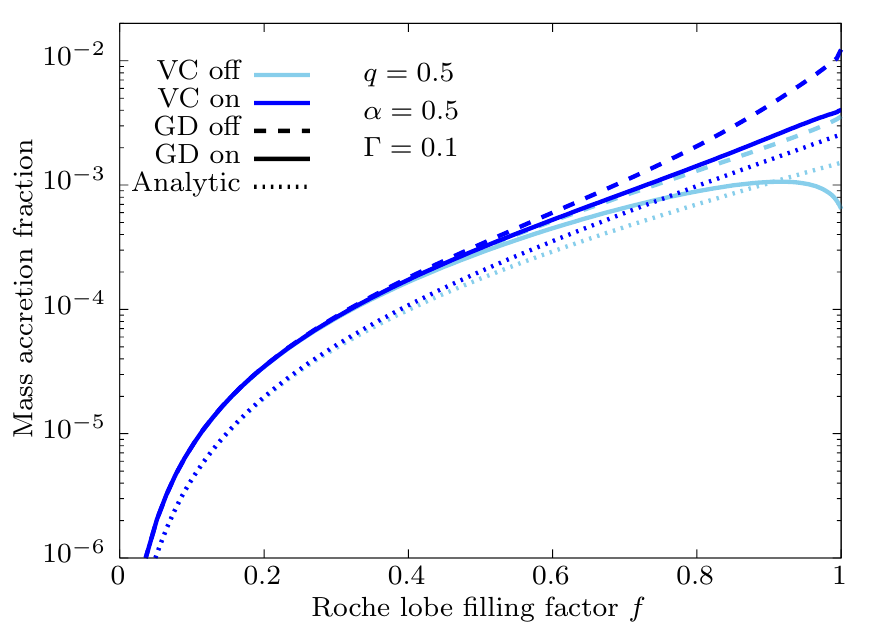}
 \caption{Mass accretion fraction as a function of Roche lobe filling factor $f$. Colours of the curves indicate whether the VC was included or not for the wind integration. Dashed curves show results without GD whereas solid curves are for models with GD. Dotted curves are analytical estimates using the BHL accretion model.\label{fig:macc_q05}}
\end{figure}

The total mass that accretes onto the secondary object is heavily influenced by the hydrodynamical processes that cannot be followed in our ballistic approach. Nevertheless, we can obtain upper limits on the mass accretion rate by integrating the mass flux through the accretion radius $r_\mathrm{acc}$. This is an upper limit because not all of this mass is guaranteed to accrete. In Figure~\ref{fig:macc_q05} we show the wind mass flux through the accretion radius normalized by the total mass-loss rate from the donor assuming a spherical wind. When the VC and GD are both switched off (light dashed curve), the mass accretion fraction monotonically increases with $f$, reaching up to $\sim 0.4~\%$ for fully Roche lobe filling cases. As we switch on the VC (dark dashed curve), the wind velocity around the equator decreases and thus the accretion radius increases, leading to higher mass accretion fractions ($\lesssim1.1~\%$). On the other hand, if we consider GD without the VC (light solid curve), the overall accretion rate decreases and the maximum accretion rate is achieved where the star is slightly underfilling its Roche lobe. In our ``full'' model with both the VC and GD switched on (dark solid curve), the mass accretion fraction is again reduced compared to the model with only the VC switched on, but remains monotonic. It is surprisingly similar to the ``vanilla'' model where both the VC and GD are switched off. The overall trend can be well described by a simple BHL accretion model (dotted curve, see Appendix~\ref{app:BHL} for details), although the absolute value is slightly larger due to the gravitational focusing. Our full model can be regarded as a conservative estimate, since the GD coefficient could be lower ($\beta_1<1$) and we have also ignored the bi-stability mechanism and X-ray feedback.

\begin{figure}
 \centering
 \includegraphics[width=\linewidth]{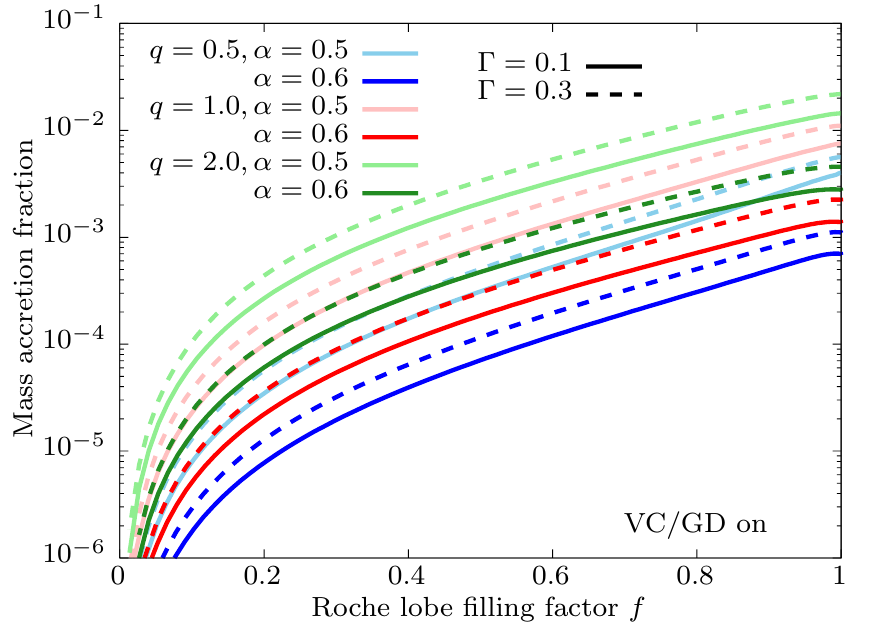}
 \caption{Mass accretion fractions for various models with different mass ratios and wind acceleration parameters. All curves were computed with both VC and GD switched on.\label{fig:macc_all}}
\end{figure}

We compare the mass accretion fractions for models with different model parameters in Figure~\ref{fig:macc_all}. The most influential parameter is the wind acceleration parameter $\alpha$, which has a direct effect on the wind terminal velocity. For larger values of $\alpha$ (darker colours), the accretion radius shrinks and thus the mass accretion fraction drops substantially. The Eddington factor also plays a role in determining the terminal velocity, where larger values of $\Gamma$ lead to lower velocities. This also affects the mass accretion fraction (dashed vs solid curves) but has a weaker effect compared to varying $\alpha$. The mass ratio dependence is straightforward. Heavier accretors (larger $q$) have larger accretion radii and therefore have greater accretion fractions.

In Figure~\ref{fig:surface_grav}c, we have circled the region on the donor star from which the accreted material originates. The whole region is displaced from the axis pointing towards the accretor by $\sim14\degree$. Most of the region is within a severely gravity darkened part of the star, which explains the stark difference in the mass accretion rate between models with GD turned on and off.

\subsection{Accreted angular momentum}

\begin{figure}
 \centering
 \includegraphics[width=\linewidth]{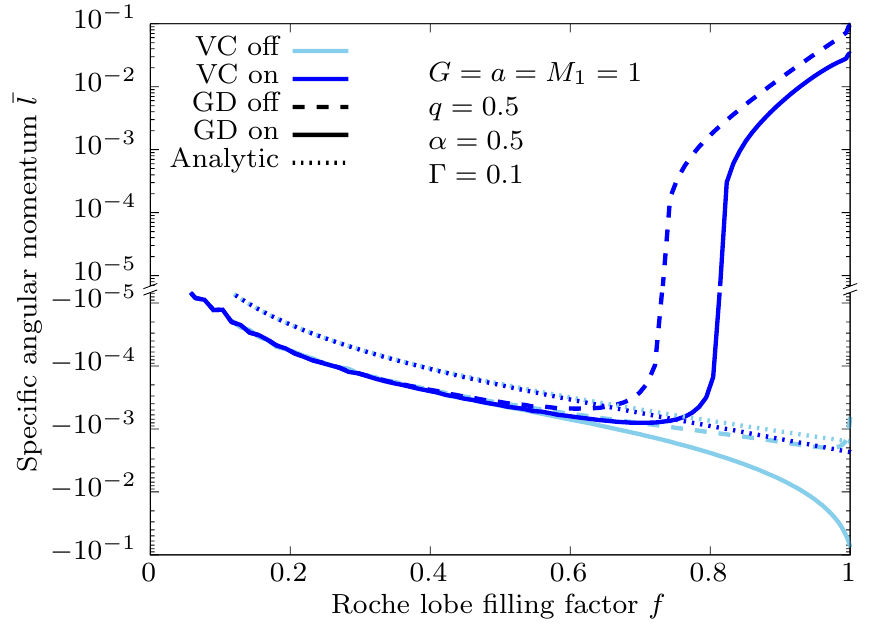}
 \caption{Angular momentum crossing the accretion radius as a function of Roche lobe filling factor $f$. Dotted curves show the analytical prediction using Eq.~(\ref{eq:analytic_AM}). Note the gap in the ordinate as angular momentum flips signs. \label{fig:AM_q05}}
\end{figure}

In Figure~\ref{fig:AM_q05} we show the average specific angular momentum $\bar{l}$ of the material crossing the accretion radius. This is computed by dividing the total angular momentum flux by the total mass flux through the accretion sphere. When the VC is switched off (light coloured curves), the angular momentum monotonically decreases with $f$, taking negative values throughout. This means that the angular momentum that accretes onto the accretor is opposite to the orbital angular momentum, which also means it is likely opposite to the spin of the accretor\footnote{The accretor would have been tidally spun up before collapsing into a compact object if the orbital separation was tight enough.}. In this case, the accretor would be spun ``down'' by the mass accretion. When the VC is switched on (dark coloured curves), the results dramatically change and show a positive angular momentum in the high-$f$ regime. This is due to the Roche-lobe overflow-like accretion stream that develops only when the VC is switched on (Figure~\ref{fig:streamlines_q05}). In this regime, the accretor would likely be spun ``up'' by the accretion, although BH spins in HMXBs are not expected to significantly evolve through accretion due to the short lifetime of the donor relative to the Eddington timescale \citep{KingKolb:1999}.

GD reduces the magnitude of specific angular momentum in the high-$f$ regions (solid curves). This is because the mass flux in the dense stream is responsible for the positive component of the angular momentum, which is severely quenched when GD is taken into account. We can understand this from Figure~\ref{fig:surface_grav}c. GD causes a steep flux gradient from the L1 point outwards and within the white circled region, there is a clear flux gradient from right to left. The left half of the region is responsible for the positive angular momentum and the right half is responsible for the negative angular momentum. The left half is more quenched, leading to the reduction in total angular momentum. However, in most cases this effect is not enough to change the qualitative behaviour where there is a transition from negative to positive values as long as the VC is taken into account.

\begin{figure}
 \centering
 \includegraphics[width=\linewidth]{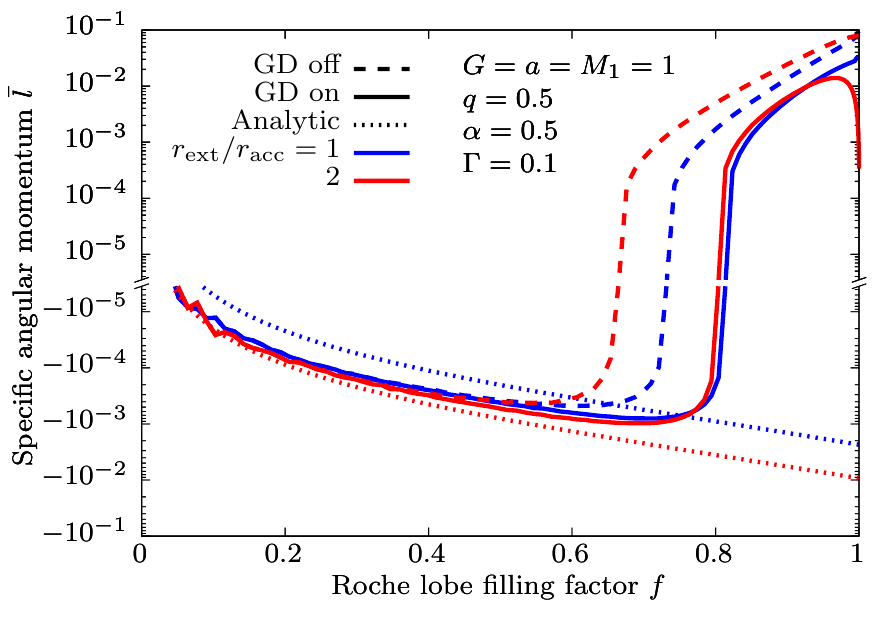}
 \caption{Comparison of angular momentum crossing different accretion radii. All models were computed with the VC on.\label{fig:racc_comparison}}
\end{figure}

The dotted curves show the analytic value estimated by the formula presented in \citet{ill75} which is expressed as
\begin{equation}
 j_\mathrm{ana}=-\frac14\Omega r_\mathrm{acc}^2.\label{eq:analytic_AM}
\end{equation}
At low filling factors, the numerically computed values are in close agreement with the shape of the analytic curve. The difference in absolute values is not trivial. In Figure~\ref{fig:racc_comparison}, we show results of the same computation with different choices for the size of the extended accretion radius. The analytic formula predicts that the intersected specific angular momentum increases as we increase the size of the accretion radius (dotted curves). However, the numerical results seem to be insensitive to the choice of accretion radius at least for the lower filling factors. The only major difference is in the transition point from positive to negative values. Also, there is a dip at $f=1$ only for the model with $r_\mathrm{ext}=2~r_\mathrm{acc}$ and GD on. The fine details sensitive to the choice of accretion radius should not be over-interpreted, but the qualitative behaviour such as the existence of a transition from a focused flow to BHL accretion remain unchanged.

\begin{figure}
 \centering
 \includegraphics[width=\linewidth]{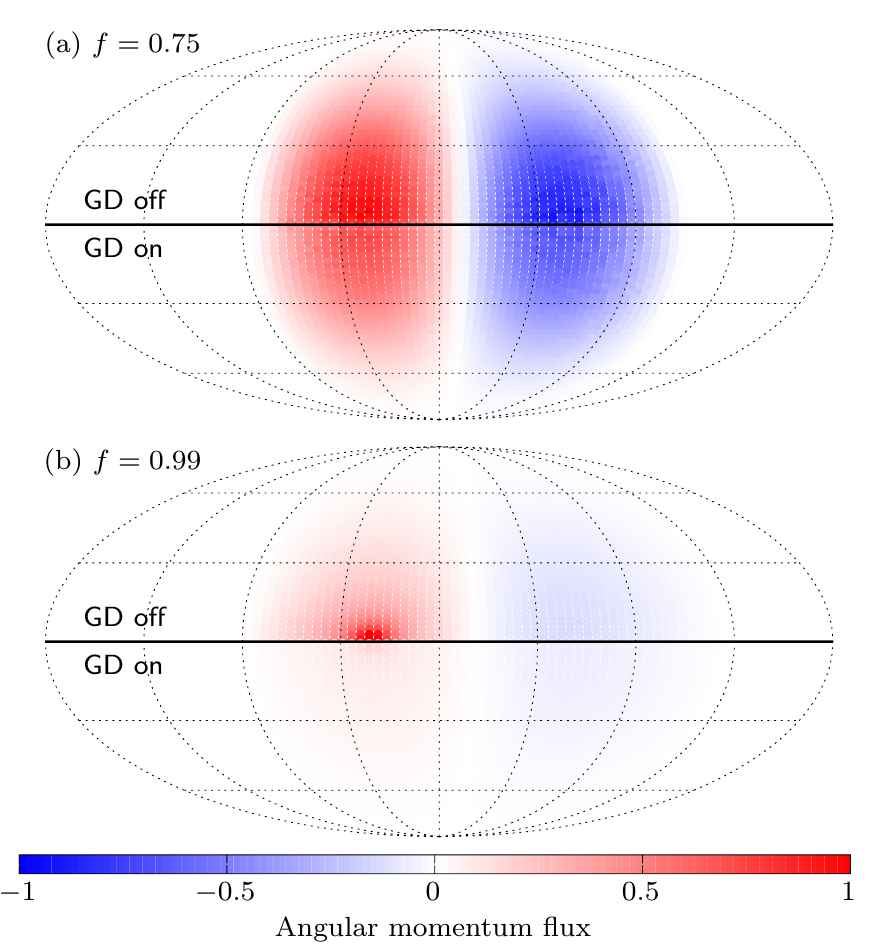}
 \caption{Angular momentum influx at the extended accretion radius from the point of view of the accretor. The centre corresponds to the direction of the donor. The top and bottom panels show results with Roche lobe filling factors $f=0.75$ and $f=0.99$, respectively. The upper halves of both panels show results with GD switched off and the lower halves shows results with GD switched on. Fluxes are normalized by the maximum value in the panel.\label{fig:influxmap}}
\end{figure}

The two qualitatively different accretion regimes can be visually identified in Figure~\ref{fig:influxmap}. For lower values of $f$ (top panel), the accretion is nearly symmetrical: material with both positive and negative angular momenta flows in and the angular momenta mostly cancels out. The effect of GD is also minor, so the upper and lower halves of panel (a) are similar. At high filling factors (bottom panel), we can clearly see a concentrated stream flowing in from one side. The net angular momentum is therefore dominated by this stream with positive values. The effect of GD is also more dramatic here, and we can see the total flux is severely quenched in the lower half of panel (b).

\begin{figure}
 \centering
 \includegraphics[width=\linewidth]{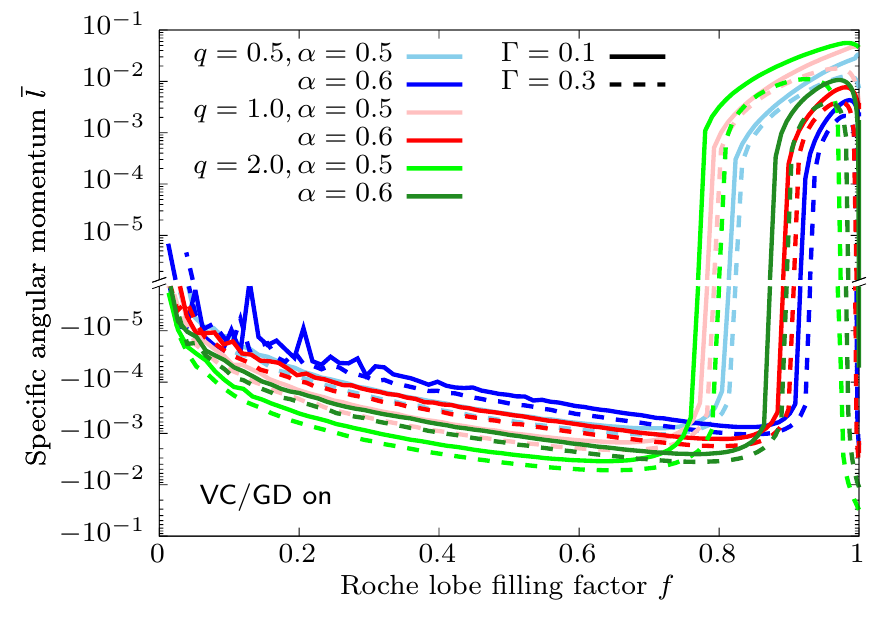}
 \caption{Parameter study of the accreted angular momentum. Colours of the curves indicate the mass ratio. Light colours are for models with $\alpha=0.5$ and dark colours for $\alpha=0.6$. Dashed curves were computed with $\Gamma=0.3$ whereas the solid curves are for $\Gamma=0.1$.\label{fig:am_all}}
\end{figure}

We summarize our results of the parameter study in Figure~\ref{fig:am_all}. The qualitative behaviours are similar for all parameter sets. All curves have negative angular momenta at the lower filling factors, roughly following BHL-like accretion. The spiky features at the low-$f$ end are due to numerical resolution issues that arise when the accretion radius is small. At the high filling factors, the angular momenta increase and reach positive values, indicating the formation of a focused stream. The transition from wind-dominated to stream-dominated accretion strongly depends on the wind acceleration parameter $\alpha$. Lower values of $\alpha$ lead to slower winds, which are therefore  more strongly gravitationally focussed. Within the parameter range we explored, the transition occurs at around $f\sim0.8$--0.9.

\section{Discussion}\label{sec:discussion}

\subsection{Disc formation}
\acp{HMXB} are known to harbour accretion discs around the compact objects, especially in the so-called high-soft state. One of the necessary conditions to create an accretion disc is for the accreting material to have enough specific angular momentum such that it exceeds that of the \ac{ISCO}. Once a disc is formed, material will slowly move inward and transport excess angular momentum outwards via viscosity, magnetic fields, etc., at the same time converting part of gravitational energy into radiation. Without sufficient angular momentum, the matter would likely fall directly onto the BH, with radiatively inefficient accretion.

The \ac{ISCO} radius for a non-rotating \ac{BH} is $r_\mathrm{isco}=3~R_\mathrm{S}$, where $R_\mathrm{S}=2GM/c^2$ is the Schwarzschild radius, but changes for spinning \acp{BH}. Hence the required angular momentum for disc formation also changes depending on the magnitude and direction of the \ac{BH} spin. For example, the required specific angular momentum for a non-rotating \ac{BH} with mass $M_\bullet=1$ is $\hat{l}\geq2\sqrt{3}\sim3.46$, whereas for a maximally spinning \ac{BH} it is $\hat{l}\geq2/\sqrt{3}\sim1.15$ for prograde orbits and $\hat{l}\leq-22/3\sqrt{3}\sim-4.23$ for retrograde orbits \citep[]{bar72}. These are all in natural units ($G=c=1$).  Note that the lower threshold on $\hat{l}$ for prograde disc formation around rapidly spinning black holes enhances the probability of detecting such systems as HMXBs \citep{sen21}.

We have used units of $G=a=M_1=1$ for all our wind calculations shown above. We can convert the specific angular momentum from our units ($G=a=M_1=1$) to units of accretor gravitational radii ($G=c=M_2=1$) through
\begin{align}
 \hat{l}&=\bar{l}\cdot \frac{2}{\sqrt{1+q}}\left(\frac{a}{R_\mathrm{S,2}}\right)\left(\frac{v_\mathrm{orb}}{c}\right)\\
 &\sim\bar{l}\cdot 730 \frac{(1+q)^{1/6}}{q}\left(\frac{P}{5.6~\mathrm{d}}\right)^\frac13\left(\frac{M_1}{40~\msun}\right)^{-\frac13},\label{eq:unit_conversion}
\end{align}
where $c$ is the speed of light and $R_\mathrm{S,2}$ is the Schwarzschild radius of the accretor. We used the orbital period and donor mass of Cyg X-1 \citep[]{mil21} as fiducial values.

For the ``full'' (VC+GD on) model in Figure~\ref{fig:AM_q05}, the accreted specific angular momentum reaches up to $\bar{l}\sim3\times10^{-2}$ at Roche lobe filling factors $f\gtrsim0.9$. Converted into appropriate units for Cyg X-1, which has a high Roche-lobe filling factor $f\sim0.997$ and mass ratio $q \sim 0.5$ \citep[]{mil21}, this yields $\hat{l}\sim46.8$, which easily exceeds the required angular momentum for disc formation for any value of the \ac{BH} spin. On the other hand, for the model without VC or GD, the angular momentum is $\bar{l}\sim-1.3\times10^{-3}$ at $f=0.997$, and therefore $\hat{l}\sim-2.0$. This is short of the threshold for retrograde disc formation around a maximally spinning black hole ($\sim-4.23$). Since the BH in Cyg X-1 is known to be rapidly spinning \citep[$\chi>0.95$][]{gou11,gou14,dur16,Zhao:2021}, this may be too small to sustain a counter-rotating accretion disc. Also, the observations support a prograde accretion disc, ruling out any retrograde accretion. This indicates that the VC could be a critical ingredient in achieving a situation where disc formation is possible. Within the parameter space we explored, the threshold for forming an accretion disc is only achieved at $f\gtrsim$0.8--0.9 (Figure~\ref{fig:am_all}). This suggests that accretion discs may only be found around \acp{BH} in systems where the donor's Roche lobe filling factor is $f\gtrsim$0.8--0.9 in HMXBs with parameters similar to those of Cyg X-1.

We caution that our results only approximately indicate the true value of the accreted angular momentum. The flow around the accretor can be strongly affected by hydrodynamical interactions between streamlines, shocks, disc winds, X-ray feedback, etc.~\citep[]{blo90,blo91,nag04,had12,cec15,elm19,mac20}. These effects are not captured in our ballistic approach. Instead, our approach can be regarded as an isothermal limit, where cooling is infinitely efficient. 3D radiation-hydrodynamic simulations are required to correctly understand the true morphology of the accretion flow, disc formation and emission properties. There can be further complications if the accretor is a neutron star with a hard surface and magnetic fields \citep[]{sha15,kar19,kar20}. We leave such exploration for future work.

\subsection{X-ray luminosity}

The strength of X-rays from \acp{HMXB} is directly proportional to the mass accretion rate, which in turn is proportional to the mass-loss rate of the donor. For a single non-rotating star with mass, temperature and luminosity similar to the donor of Cyg X-1, this is $\dot{M}\sim2\times10^{-6}~\msun~$yr$^{-1}$ (see Appendix~\ref{app:mdot} for details). This value corresponds to the polar value, which is not sensitive to VC/GD. In Figures~\ref{fig:macc_q05} and \ref{fig:macc_all}, we defined the mass accretion fraction relative to the polar mass flux. Therefore, in order to estimate the mass accretion rate onto the compact object, we simply need to multiply the fraction by the polar mass flux.

For Cyg X-1, the mass ratio is $q\sim0.5$ \citep[]{mil21} so we can use Figure~\ref{fig:macc_q05} to estimate the mass accretion rate. The mass accretion fraction at $f\sim0.997$ is about $\sim0.004$, so if we adopt $\dot{M}=2\times10^{-6}~\msun$~yr$^{-1}$ as the polar mass flux, the mass accretion rate is roughly $\dot{M}_\mathrm{acc}\sim8\times10^{-9}~\msun$~yr$^{-1}$. This translates to an X-ray luminosity from the accretor of
\begin{align}
 L_X&=\epsilon\dot{M}_\mathrm{acc}c^2\nonumber\\
&\sim4.5\times10^{37}\mathrm{erg~s}^{-1}\left(\frac{\epsilon}{0.1}\right)\left(\frac{\dot{M}_\mathrm{acc}}{8\times10^{-9}~\msun~\mathrm{yr}^{-1}}\right),
\end{align}
where $\epsilon$ is the radiative efficiency of accretion. This is in rough agreement with observed X-ray luminosities from Cyg X-1 \citep[$\sim$1--3$\times10^{37}$~erg~s$^{-1}$;][]{vrt08,sug17}.

\subsection{Implications}

When the accretion flow cannot form a disc, we expect the radiative efficiency to be much lower as the wind material will directly fall into the BH without converting its gravitational energy into radiation. If so, HMXBs with and without discs should have drastically different X-ray luminosities. As the donor stars in binaries evolve and grow in size, the Roche lobe filling factor will gradually increase. Soon after the transition point from negative to positive accreted angular momentum (e.g. Figure~\ref{fig:am_all}), the system should suddenly switch on as an HMXB as it exceeds the threshold for disc formation. This is a much steeper transition than in the simple BHL picture where the X-ray luminosity is assumed to be directly proportional to the mass-accretion rate. It thus imposes a strong observational bias against systems with $f\lesssim0.8$. Of course, there are many effects that could blur this transition such as the instability and clumpiness of line-driven winds, stellar variability, etc. Nevertheless, it is interesting that all the wind-fed HMXBs with confirmed BHs have donors with Roche lobe filling factors $f\gtrsim0.8$ (see Section~\ref{sec:introduction}). 

Despite the simplifying assumptions in our model (see Section~\ref{sec:caveats}), the Roche-lobe filling factor limit is qualitatively robust. X-ray irradiation from the accretor is known to have profound effects on the wind driving and therefore the X-ray luminosity \citep[e.g.][]{hat77}. However, this only occurs when there is sufficient X-ray emission from the accretor in the first place. Our results show that when the Roche lobe filling factor is below $f\lesssim0.8$, accretion cannot generate X-rays efficiently, so should be unable to trigger the X-ray irradiation effect. Therefore, we expect the transition threshold from X-ray dim to X-ray bright binaries at around $f\gtrsim$0.8--0.9 to be rather robust. In fact, the self-regulation through X-ray irradiation discussed in Section~\ref{sec:caveats} may make this transition even steeper. We stress that the transition is only possible with the VC switched on within our models.

It would be interesting to extend our model to the case of eccentric systems. In eccentric systems, it is not possible for the donor to be tidally synchronised to the orbit. Therefore, the degree of gravity darkening may be much more limited. On the other hand, the surface gravity distribution will nevertheless be non-uniform, especially around periastron, so the wind velocity field may be altered in a similar way to our models. This could lead to a phase-dependent wind field which may have interesting observational consequences. However, this depends on how the wind driving reacts to a time-dependent gravitational field, which may not be the same as the solution for stationary states that we assume in this paper. Such states would not last long though, because in tight binaries the orbit will be quickly tidally circularized. We leave this investigation for future work.

\section{Conclusion}\label{sec:conclusion}
We suggest that line-driven winds can be quite anisotropic in tight X-ray binary systems with circular orbits and tidally locked donor stars. We focus on two effects that can influence the wind driving. One is the reduction in wind terminal velocity due to rotation. This has been discussed in the context of single rotating stars such as Be stars, but not so much in the context of binaries. In tight X-ray binaries, the donor star is usually tidally locked to the orbit, achieving rapid rotation. Therefore the wind velocity should be reduced around the equator. In addition, the tidal field due to the presence of the companion creates an inhomogeneous surface gravity distribution on the donor. This may further cause the wind velocity field to be non-axisymmetrical.

The other effect is gravity darkening. Tidal fields in close binaries are known to create non-uniform flux distributions across the surface of the star due to von Zeipel's theorem. This is usually photometrically identified as the ellipsoidal variation. Given that stellar winds are driven by radiation, the gravity darkened regions should have lower mass flux. 

We modelled the possible anisotropic wind structure in X-ray binaries based on the CAK theory for line-driven winds. We created a simple way to account for the effect of velocity reduction and gravity darkening by appropriately scaling the wind acceleration and mass flux through line driving. Using this simple method, we calculated the possible amount of mass and angular momentum that would accrete onto the compact object.

With the velocity correction alone, the total accreted mass increases due to the slower wind velocities towards the accretor. However, once both the velocity correction and gravity darkening effects are accounted for, the total accreted mass is similar to the simple Bondi-Hoyle-Lyttleton accretion model. 

We identify the formation of a focused stream reminiscent of ``wind Roche lobe overflow'' when the velocity correction is taken into account. This significantly enhances the amount of accreted angular momentum, enabling the formation of an accretion disc. Since such focused streams and accretion discs are known to exist in observed X-ray binaries, it may imply that the velocity correction is playing a role in the wind driving. Our results suggest the threshold value of the Roche lobe filling factor to create this focused stream is around $f\sim$0.8--0.9, depending on the wind force parameters. Our model is only applicable for wind-fed high-mass X-ray binaries with circular orbits and tidally locked OB-type donor stars. We predict that within such systems, none of them with Roche lobe filling factors $f<0.8$ harbour accretion discs. Therefore, there should be a discrete step difference in the observability of high-mass X-ray binaries depending on whether the Roche lobe filling factor exceeds this value or not.

High Roche lobe filling factors can be achieved as the donor star grows in size due to stellar evolution. \citet{van20} and \citet{sen21} showed that some of the known Galactic Wolf-Rayet+O-star binaries will eventually evolve into BH+O-star binaries with high Roche lobe filling factors.  They find that high Roche lobe filling factors reduce the wind velocity, increase the mass accretion rate, and allow the BH HMXBs to be observed.  Our models, which include the effects of rotation and gravity darkening,
are consistent with this finding and additionally indicate that these observable systems will have prograde accretion instead of retrograde accretion that is naturally expected in wind-fed systems.

\begin{acknowledgements}
The authors thank Koushik Sen, Xiao-Tian Xu and Ileyk El Mellah for insightful comments on our preliminary draft. This work was performed on the OzSTAR national facility at Swinburne University of Technology. The OzSTAR program receives funding in part from the Astronomy National Collaborative Research Infrastructure Strategy (NCRIS) allocation provided by the Australian Government.
IM is a recipient of the Australian Research Council Future Fellowship FT190100574.
\end{acknowledgements}

\begin{appendix}
\section{Accretion fraction in the simple BHL model}\label{app:BHL}
In the simple BHL model, we assume that the wind is spherically symmetric, with the same velocity distribution and mass flux along each direction. In this situation, the accretion fraction $f_\mathrm{acc}$ can be obtained by simply calculating the solid angle of the accretion sphere as viewed from the centre of the donor. This gives
\begin{equation}
 f_\mathrm{acc}=\frac{1-\sqrt{1-(r_\mathrm{acc}/a)^2}}{2}\sim \frac14\left(\frac{r_\mathrm{acc}}{a}\right)^2,
\end{equation}
where $r_\mathrm{acc}$ is the accretion radius and $a$ is the orbital separation. The ratio between accretion radius and separation can be expressed as 
\begin{equation}
 \frac{r_\mathrm{acc}}{a}=\frac{2GM_2}{v_\bullet^2a}=\frac{2q}{1+q}\left[1+\left(\frac{v_\mathrm{w}}{v_\mathrm{orb}}\right)^2\right]^{-1},
\end{equation}
where $v_\mathrm{w}$ is the wind velocity at the position of the accretor and $v_\mathrm{orb}$ is the orbital velocity. For the wind velocity, we use $v_\mathrm{w}=v(a)$ based on Eq.~(\ref{eq:vwind_fd_beta}). The stellar radius $R$ can be related to the Roche lobe filling factor through $R/a=f\mathcal{E}(q)$ where 
\begin{equation}
 \mathcal{E}(q)\equiv\frac{0.49q^{-2/3}}{0.6q^{-2/3}+\ln(1+q^{-1/3})},
\end{equation}
is the ratio between the Roche lobe radius and orbital separation \citep[]{egg83}. Combining these yields
\begin{equation}
 \frac{v_\mathrm{w}}{v_\mathrm{orb}}=2.5 \frac{\alpha}{1-\alpha}\left(1-f\mathcal{E}(q)\right)^{0.7}\sqrt{\frac{2(1-\Gamma)}{(1+q)f\mathcal{E}(q)}}.
\end{equation}

\section{Mass-loss rates from CAK theory}\label{app:mdot}
\begin{figure}
 \centering
 \includegraphics[width=\linewidth]{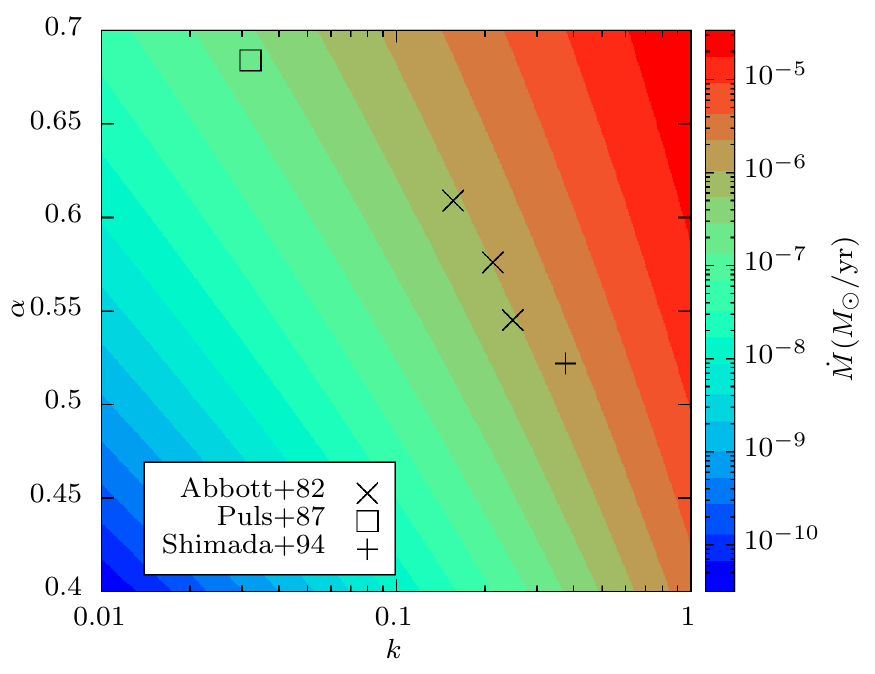}
 \caption{Mass-loss rate predicted from CAK theory using Eq.~(\ref{eq:mdot_CAK}). The stellar parameters are set to those of Cyg X-1 \citep[$M=40~\msun$, $\Gamma=0.2$, $T_\mathrm{eff}=$31,000 K;][]{mil21}. Black symbols indicate the force parameters computed for $T_\mathrm{eff}\sim$30,000 K atmospheres in \citet{abb82,pul87,shi94}.\label{fig:mdot_CAK}}
\end{figure}

In CAK theory, the radiative acceleration is nonlinearly coupled to density, so the mass-loss rate is uniquely determined given the critical solution for velocity \citep[]{cas75}. By equating Eq.~(\ref{eq:Cdef}) with Eq.~(\ref{eq:C_standard}), we get
\begin{equation}
 \dot{M}=\frac{k^\frac{1}{\alpha}}{v_\mathrm{th}}\frac{L_\mathrm{Edd}}{c}\Gamma^\frac{1}{\alpha}(1-\Gamma)^{1-\frac{1}{\alpha}}\alpha(1-\alpha)^{\frac{1}{\alpha}-1}(1+\alpha)^{-\frac{1}{\alpha}},\label{eq:mdot_CAK}
\end{equation}
where $L_\mathrm{Edd}\equiv4\pi GMc/\sigma_e$ is the Eddington luminosity. We plot this function in Figure~\ref{fig:mdot_CAK}, with stellar parameters chosen to match the donor of Cyg X-1. The mass-loss rate can range from $<10^{-10}~\msun$~yr$^{-1}$ to $>10^5~\msun$~yr$^{-1}$ within the range of possible force multiplier parameters. However, $k$ and $\alpha$ are usually correlated and are not independent parameters \citep[]{gay95}. The black symbols indicate combinations of force parameters computed for stellar atmospheres with $T_\mathrm{eff}\sim$30,000 K \citep[]{abb82,pul87,shi94}. Despite the large spread in the values of $k$ and $\alpha$, the resulting mass-loss rates have similar values around $\sim10^{-6}~\msun$~yr$^{-1}$. This is in rough agreement with the mass-loss rate of the Cyg X-1 donor inferred from H$\alpha$ line measurements \citep[$\sim$2--5$\times10^{-6}~\msun$~yr$^{-1}$][]{her95,gie03,vrt08}.

\end{appendix}

\bibliographystyle{pasa-mnras}
\providecommand{\NOOPSORT}[1]{}

\end{document}